\documentclass[english,aps,preprint,5p]{elsarticle}
\usepackage[T1]{fontenc}
\usepackage[latin9]{inputenc}
\usepackage{float}
\usepackage{units}
\usepackage{textcomp}
\usepackage{relsize}
\usepackage{graphicx}
\usepackage{color}
\usepackage{booktabs}
\usepackage{tabularx}
\usepackage{wasysym}
\makeatletter

\@ifundefined{showcaptionsetup}{}{%
 \PassOptionsToPackage{caption=false}{subfig}}
\usepackage{subfig}
\makeatother

\usepackage{babel}

\begin{document}

\title{Fabrication of Diamond Nanowires for Quantum Information Processing Applications}

\author[harv,mun]{Birgit J. M. Hausmann}
\ead{hausmann@seas.harvard.edu}
\author[harv]{Mughees Khan}
\author[harv]{Yinan Zhang}
\author[harv]{Tom M. Babinec}
\author[newmex]{Katie Martinick}
\author[harv]{Murray McCutcheon}
\author[texas]{Phil R. Hemmer}
\author[harv]{Marko Lon$\mathrm{\breve{c}}$ar}

\address[harv]{School of Engineering and Applied Sciences, Harvard University, McKay Lab 219, 9 Oxford Street, Cambridge MA, 02138, United States}
\address[mun]{Department of Physics, Technische Universität München, James Franck Str., D-85748 Garching, Germany}
\address[newmex]{Department of Chemical and Nuclear Engineering, The University of New Mexico, 1 University of New Mexico, Albuquerque, NM 87131-0001, United States}
\address[texas]{Department of Electrical and Computer Engineering, Texas A\&M University, 216H Zachry, TAMU 3128 College Station, Texas 77843-3128, United States}

\cortext[cor1]{Corresponding author: Tel.: +1 617 2300238}

\begin{abstract}
We present a design and a top-down fabrication method for realizing diamond nanowires in both bulk single crystal and polycrystalline diamond. Numerical modeling was used to study coupling between a Nitrogen Vacancy (NV) color center and optical modes of a nanowire, and to find an optimal range of nanowire diameters that allows for large collection efficiency of emitted photons. Inductively coupled plasma (ICP) reactive ion etching (RIE) with oxygen is used to fabricate the nanowires. Drop-casted nanoparticles (including $\mathrm{Au}$, $\mathrm{SiO_{2}}$ and $\mathrm{Al_2O_3}$) as well as electron beam lithography defined spin-on glass and evaporated $\mathrm{Au}$ have been used as an etch mask. We found $\mathrm{Al_2O_3}$ nanoparticles to be the most etch resistant. At the same time FOx e-beam resist (spin-on glass) proved to be a suitable etch mask for fabrication of ordered arrays of diamond nanowires. We were able to obtain nanowires with near vertical sidewalls in both polycrystalline and single crystal diamond. The heights and diameters of the polycrystalline nanowires presented in this paper are $\unit[\approx1]{\mu m}$ and $\unit[120-340]{nm}$, respectively, having a $\unit[200]{nm/min}$ etch rate. In the case of single crystal diamond (types Ib and IIa) nanowires the height and diameter for different diamonds and masks shown in this paper were $\unit[1-2.4]{\mu m}$ and $\unit[120-490]{nm}$ with etch rates between $\unit[190-240]{nm/min}$.
\begin{keyword}
Diamond \sep Nanowire \sep Reactive ion etching \sep ICP \sep Nanostructure
\end{keyword}
\end{abstract}
\maketitle

\section{Introduction}

Nitrogen-vacancy (NV) color centers embedded in diamond have recently emerged as a promising platform for the realization of robust, room temperature, single-photon sources \cite{Weinfurtner}. The NV center is formed by a two point defect in the diamond lattice: A substitutional nitrogen atom and a vacancy (missing carbon atom) trapped at an adjacent lattice position. It can exist in two charged states $\mathrm{NV^{0}}$ and $\mathrm{NV^{-}}$ out of which the $\mathrm{NV^{-}}$ center is suitable for quantum operations due to the electron's spin kinetics. The luminescence spectrum of an $\mathrm{NV^{-}}$ color center consists of a broad ($\unit[640]{nm}$ to $\unit[780]{nm}$) phonon side-band (PSB) and a zero-phonon line (ZPL) at $\unit[637]{nm}$ \cite{Manson, Jelezko}. NV centers combine the key advantages of isolated atomic systems with solid-state integration, and have the following unique optical properties: (i) room temperature operation, (ii) little inhomogeneous broadening and (iii) deterministic positioning using ion implantation. Moreover, the NV center stands out among solid state systems because its electronic spin can be  prepared, manipulated, and measured with optical and microwave excitations \cite{Wrachtrup}.

However, NV centers also interact with their environment leading to decoherence in the presence of nearby nitrogen spins. Hence it is important to enhance the NV center emission yield (photon production rate) as well as increase the collection efficiency of emitted photons. This can be achieved by embedding NV centers within optical structures including cavities, microdisks, waveguides and nanowires.

$\mathrm{NV^{-}}$ optical transition is in the visible ($\unit[637]{nm}$), which limits the choice of materials that can be used to realize optical structures. Diamond itself is a suitable platform having a reasonably high refractive index ($n$ = 2.43). Baldwin et al. \cite{Houston2} have fabricated 2D photonic crystal (PhC) slabs in polycrystalline diamond (poly-D) on insulator (DOI) films. Similarly, Wang et al. \cite{Wang2, Wang4} have demonstrated the fabrication of optical microdisk resonators and 2D PhC in polycrystalline diamond films using E-Beam Lithography (EBL) and Reactive Ion Etching (RIE). The characterization of poly-D microdisks and photonic crystal cavities has shown scattering losses due to the polycrystalline nature of the diamond film. Moreover, the emission from polycrystalline diamond itself can obscure single-photon emission from an embedded NV center. Hence, defect-free single crystal diamond (sc-D) films are needed for Quantum Information Processing (QIP) applications.

High-quality sc-D films can only be homo-epitaxially grown on the diamond substrate and therefore the needed index contrast for the realization of nanophotonic devices does not exist. Olivero et al. \cite{Prawer2} have used ion-beam implantation followed by sacrificial removal of implanted graphitized (damaged) layer to fabricate thin sc-D films. Based on this approach fabrication of sc-D waveguides \cite{Prawer} and microdisk resonators \cite{Wang3} has been demonstrated. However, the ion implantation may create additional NV centers and may render such thin films unsuitable for QIP applications.

Optical structures fabricated in a GaP layer bonded on top of a diamond film \cite{Beausoleil} have also been reported. Approaches based on diamond nanocrystals containing NV centers embedded within a silicon-nitride photonic crystal structure has been examined theoretically \cite{Murray}. Other works include positioning diamond nanocrystals on top of wide bandgap materials such as GaP \cite{Jela} and silicon-dioxide \cite{Barclay} as well as metallic nanostructures \cite{Benson}.

In this work an alternative approach has been explored based on the fabrication of nanowires in diamond containing NV centers. Recently, Friedler et al. \cite{Lalanne} have theoretically demonstrated a $90\%$ extraction efficiency of emitted photons from single-photon sources like quantum dots embedded within semiconductor nanowires. We first examine an optimal nanowire diameter that maximizes the overall collection efficiency of photons emitted from an NV center. Next, we developed a fabrication procedure for the creation of diamond nanowires directly in high-quality sc-D substrates utilizing RIE and as needed EBL. Our procedure uses oxygen as a reactive gas in an inductively coupled plasma (ICP) RIE. In the past a two step procedure for diamond nano-rods based on hydrogen treated sc-D whiskers has been pursued \cite{Sawabe}. Other fabricated structures in poly- and sc-D based on reactive ion etching (RIE) include microlenses \cite{Scarsbrook, Gurney}, porous diamond \cite{Shiomi}, field emitters \cite{Fujimori2}, waveguides \cite{Prawer}, atomic force microscopy (AFM) tips \cite{Yamada}, disks \cite{Wang3}, cones and cylinders \cite{Fujimori1}, nanopillars \cite{Wang}, microcylinders \cite{Oura} and resonator arrays for RF signal processing \cite{Houston}. As masks, we investigated two main approaches: (i) nanoparticles deposited using drop-casting ($\mathrm{Au}$, $\mathrm{SiO_{2}}$, $\mathrm{Al_2O_3}$), and (ii) masks defined by EBL using positive as well as negative resist. We also investigated two types of substrates: (i) poly-D films on insulator, and (ii) sc-D diamonds grown via high-pressure high-temperature (HPHT) and chemical vapor deposition (CVD) processes.

\section{Design and Modeling} \label{sec:Design}
\begin{figure}[htbp]
\centering
\includegraphics[width=1\columnwidth]{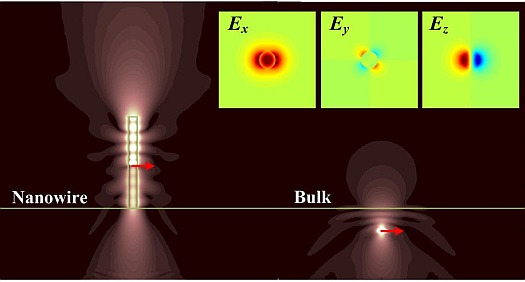}
\caption{Field profile of the E-field's radial component ($\mathrm{E_r}$) in the case of a $\unit[2]{\mu m}$ long, $\unit[200]{nm}$ diameter diamond nanowire (left) and a bulk diamond (right). The dipole is polarized parallel to the interface, emitting at $\lambda=\unit[637]{nm}$ (zero-phonon line wavelength of NV center). The fundamental nanowire waveguide mode ($\mathrm{HE_{11}}$) (inset) is the dominant decay channel for the nanowire case.
}
\label{fig:neff}
\end{figure}

\begin{figure*}[htbp]
\centering
\includegraphics[width=1\textwidth]{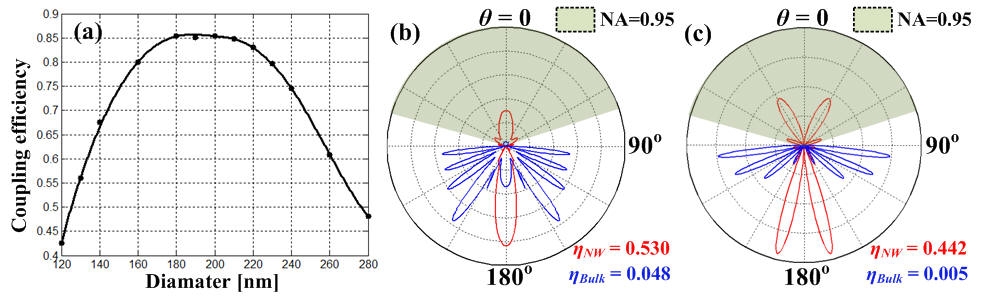}
\caption{(a) Coupling efficiency $\alpha$ as a function of the nanowire diameter at $\lambda = \unit[637]{nm}$. The dipole is assumed to be placed at the center of the nanowire with a polarization perpendicular to the nanowire axis (s-polarized). (b) and (c): The far-field profile of power emitted from a (b) s-polarized and (c) p-polarized (parallel to the nanowire axis) dipole embedded at the center of a nanowire with $d=\unit[200]{nm}$. Blue lines, in both (b) and (c), show far-field profiles of emitter embedded in the bulk diamond crystal. The shaded areas denote the light that can be collected using an objective lens with an NA of $0.95$. Calculated values of collection efficiency $\eta$ for nanowire and bulk, for both polarizations, are also indicated.}
\label{fig:purc}
\end{figure*}

First, we studied the emission from an NV center (modeled as a dipole) in a diamond nanowire using FDTD simulations. We assume a circular cross section of the nanowire with a diameter $d$, and consider the two general polarization scenarios for a dipole/nanowire system: dipole polarization perpendicular (s-polarized) and parallel (p-polarized) to the nanaowire axis. The dipole associated with an NV center in (100) diamond can be represented using combination of these two dipoles, since it is polarized in the (111) plane. At different wavelengths within the NV center's radiation spectrum ($\unit[637]{nm}-\unit[780]{nm}$), the number of collected photons per second can be expressed as $\Gamma(\lambda)\cdot\eta(\lambda)$ where $\Gamma$ is the emission rate (reciprocal to the lifetime) and $\eta$ the collection efficiency.
The collection efficiency can be dramatically improved in diamond nanowires compared with bulk diamond, as demonstrated in Figure \ref{fig:neff}. These field profiles show that the major portion of light emitted from an NV center in bulk diamond leaks to the substrate due to significant total internal reflection at the diamond-air interface, whereas in a diamond nanowire the fundamental $\mathrm{HE_{11}}$ mode is the dominant emission channel for a dipole polarized perpendicular to the nanowire axis (in the $xy$ plane) (Fig. \ref{fig:neff}, inset) \cite{Lalanne}. This waveguide mode directs the light propagating in the nanowire, and is scattered vertically as it exits from the top nanowire facet. This process allows for efficient collection using an objective lens positioned above \cite{Ning}.
In Figure \ref{fig:purc}a we show the coupling efficiency, $\alpha$, between the NV center and the nanowire waveguide mode as a function of the nanowire diameter, for wavelength $\lambda = 637nm$. It can be seen that, in the case of s-polarized dipole, more than 80\% of emitted photons can couple to the nanowire mode for a broad range of nanowire diameters ($\unit[180]{nm}-\unit[230]{nm}$). We choose to work with $\unit[200]{nm}$ diameter nanowires in order to optimize the coupling efficiency. Photon collection efficiencies can be quantified from the far-field profile of power emitted upward, shown in Figure \ref{fig:purc}b and c. An objective lens with a numerical aperture NA=0.95, positioned above the nanowire, can collect light emitted into the solid angle of 72$\mathrm{\,^{\circ}}$ (represented by shaded areas in the far field emission profiles). It can be seen that in the case of both s- and p-polarized dipoles, almost ~100\% of photons emitted from the nanowire can be collected with the lens. It is interesting to note that this is true even for p-polarized dipole despite the fact that it cannot couple to the nanowire waveguide mode due to the symmetry mismatch. In this case, however, large collection efficiency is enabled by coupling to radiative modes that are also modified by the presence of the nanowire. Finally, comparing a dipole in a nanowire (red line in figures \ref{fig:purc}b and c) with a dipole in a bulk diamond crystal (blue line), we find that the nanowire geometry provides one and two orders of magnitude improvement in the collection efficiency in the case of s-polarized and p-polarized dipole, respectively.
\begin{figure*}[htbp]
\begin{centering}
\includegraphics[width=0.8\textwidth]{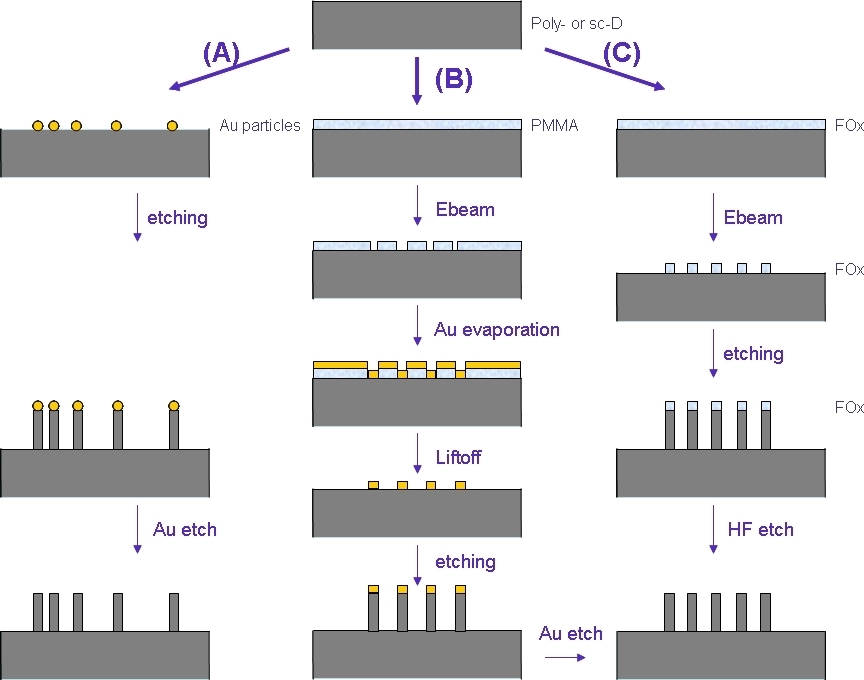}
\par\end{centering}
\caption{Fabrication process schematic. (A): Drop casting of nanoparticles ($\mathrm{Al_2O_3}$, $\mathrm{Au}$, $\mathrm{SiO_2}$) followed by etching. (B) EBL defined $\mathrm{Au}$ evaporation and etching. (C) EBL defined FOx  mask (flowable oxide, a spin-on-glass resist) and etching.}
\label{fig:cartoon}
\end{figure*}
We also evaluated the total emission rate (the reciprocal of lifetime) of an NV center in a nanowire, and found that it is dependent on the NV's position: Fabry-Perot resonances, formed due to the (weak) reflection of a waveguide mode from the nanowire's facets, can modify the emission rate of s-polarized dipole. We introduce the enhancement factor $E(\lambda)=\frac{\Gamma(\lambda)}{\Gamma_0(\lambda)}$ where $\Gamma_0$ is the emission rate of the quantum emitter in a homogeneous diamond medium. For $\lambda = 637$nm and $d = 200$nm, the enhancement factor is in the range of $0.65-1.10$, depending on the dipole position along the nanowire positions. It is interesting to note that the collection efficiency for s-polarized dipole is maximized (Figure \ref{fig:purc}b) when destructive interference occurs between downward emitted photons and photons reflected from the top nanowire facet (Figure \ref{fig:neff}). This case favors upward emission at the expense of slightly increased radiative lifetime.

In order to estimate the total number of photons that can be collected from a nanowire geometry, it is important to take into account both s- and p-polarized components of NV dipole, as well as its broad-band emission due to the phonon sideband. The total number of collected photons is obtained by averaging over wavelengths and polarizations, by using following (unitless) figure of merit

\[Z=\frac{\int\int E(\lambda, \sigma)\eta(\lambda, \sigma)\Gamma_0(\lambda)d\sigma d\lambda}{2\pi\int\Gamma_0(\lambda)d\lambda}\] where $\sigma$ denotes the polarization angle.

Using this figure of merit, we find that the nanowire geometry provides an order of magnitude improvement ($Z\approx30\%$) over the bulk diamond case ($Z\approx3.3\%$).

\section{Materials and Methods}\label{matandmet}
We used two types of diamond samples based on their crystalline nature: a) poly-D films and b) type Ib and IIa sc-Ds synthesized via HPHT and CVD processes respectively. While sc-Ds are needed to realize single-photon sources, poly-d films have been used to optimize the etch recipe due to their low-cost and availability in large quantities.

The poly-D samples were diamond on insulator (DOI) AQUA 25 wafers ($\unit[2]{\mu m}$ poly-D film on $\unit[1]{\mu m}$ thermal $\mathrm{SiO_{2}}$ on a $\mathrm{Si}$ substrate) from Advanced Diamond Technologies, Inc. The sample size was about $\unit[3-5]{mm}$ x $\unit[3-5]{mm}$. Prior to mask deposition and reactive ion etching all poly-D samples were solvent cleaned.
The nitrogen rich HPHT Ib and very pure (nitrogen content < $\unit[0.1]{ppm}$) CVD IIa sc-Ds were obtained from Element6 (E6), and were $\unit[3]{mm}$ x $\unit[3]{mm}$ in size. We also used a CVD sc-D from Apollo Diamond, Inc. All sc-D samples had a <100> crystal orientation and prior to mask deposition, and etching all sc-Ds were cleaned for about $\unit[45]{min}$ in a boiling 1:1:1 (Nitric : Perchloric : Sulfuric) acid bath. This aggressive etch was not used with the poly-D samples as it was damaging the DOI film.

Three major types of etch masks have been explored in this work, each of which required a slightly different fabrication approach. The adopted fabrication processes are shown in Figure \ref{fig:cartoon}. The etch masks were prepared as discussed below.

\begin{enumerate}
\item Nanoparticle mask, deposited via drop-casting: In this approach drop casted nanoparticles have been used as an etch mask. $\mathrm{Au}$ colloids suspended in DI water were obtained from Ted Pella, Inc. The nominal particle sizes were $\unit[100]{nm}$ and $\unit[250]{nm}$. However, our measurements indicated a large variation with mean particle diameters of $\unit[\approx140]{nm}$ and $\unit[\approx350]{nm}$ respectively. The concentrations were $5.6\times10^{9}$ particles/mL and $3.6\times10^{8}$ particles/mL for the $\unit[100]{nm}$ and $\unit[250]{nm}$ colloids respectively. $\mathrm{Au}$ nanoparticles were further diluted in 1:1 and 1:2 ratio with DI water before dispersing on the samples. A suspension of $\mathrm{SiO_{2}}$ nanoparticles (Corpuscular, Inc.) with a diameter of $\unit[\approx213]{nm}$ and a 5\% particle concentration in a $\unit[10]{mL}$ solution was also used. The $\mathrm{SiO_{2}}$ particle-size variation was smaller than 8\%. Finally, $\mathrm{Al_2O_3}$ powder from Alfa Aesar, Inc. with $\unit[200]{nm}$ and $\unit[40-50]{nm}$ diameters were used.

\item $\mathrm{Au}$ mask defined via a lift-off process: In this approach, evaporated Au was used as an etch mask as illustrated in Figure \ref{fig:cartoon}. Bilayer PMMA from MicroChem Corp was used to define an etch pattern using a $\unit[100]{kV}$ Elionix e-beam tool. Dosages were varied from $\unit[800]{{\mu C}/{cm^2}}$ to $\unit[1600]{{\mu C}/{cm^2}}$. Next, $\unit[10]{nm}$ $\mathrm{Cr}$ and $\unit[200]{nm}$ $\mathrm{Au}$ evaporation followed by a lift-off process was used to define a metal etch mask.

\item Spin-on-glass (HSQ) mask defined by EBL:  Flowable oxide (FOx 17) from Dow Corning was diluted with MIBK in 1:2 and 1:1 ratios. Dosages from $\unit[4800]{{\mu C}/{cm^2}}$ to $\unit[8000]{{\mu C}/{cm^2}}$ were used to expose FOx resist.
\end{enumerate}

An ICP RIE system (UNAXIS shuttleline) was used to transfer the mask pattern into all the diamond samples. An etch recipe having $\unit[30]{sccm}$ oxygen flow rate, $\unit[100]{W}$ Bias power, $\unit[700]{W}$ ICP power at a $\unit[10]{mTorr}$ chamber pressure was used and will be discussed in the next section.
After RIE, gold (type TFA, Transene) and chrome etchant (type 1020, Transene) were used in approaches (A) and (B) to remove the remaining Au/Cr mask. In case (C), an HF (49\%, aq) wet etch was performed to remove the remaining FOx mask.

All images were taken with a Zeiss SUPRA field emission scanning electron microscope (FESEM) under a 86° tilt unless noted otherwise. The secondary electron (SE) detector and an acceleration voltage of $\unit[5]{kV}$ was used.

\section{Results and discussion}
\subsection{Etch recipe variation}
\begin{figure*}[bthp]
\centering
$\begin{array}{c@{\hspace{0.5cm}}c}
\includegraphics[width=0.45\textwidth]{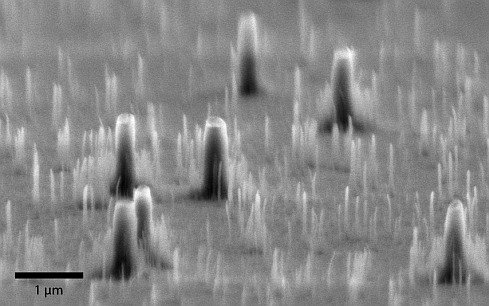} &
\hspace{0cm}\includegraphics[width=0.45\textwidth]{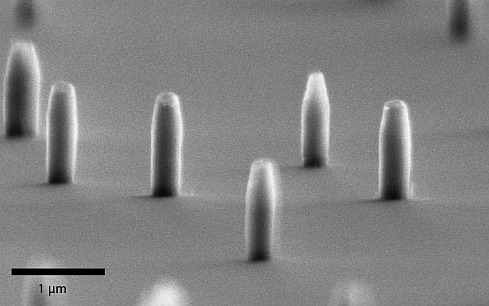} \\
(a) & (b)\\
\includegraphics[width=0.45\textwidth]{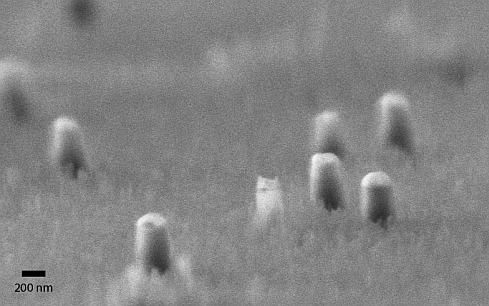} &
\hspace{0cm}\includegraphics[width=0.45\textwidth]{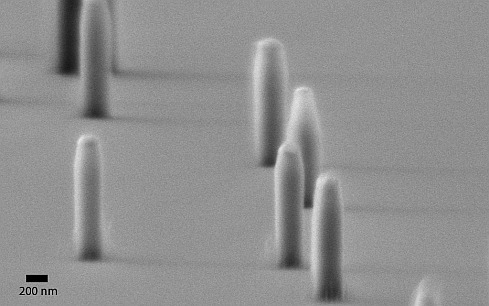} \\
(c) & (d)\\
\includegraphics[width=0.45\textwidth]{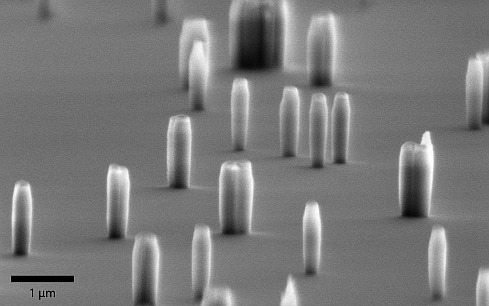} &
\hspace{0cm}\includegraphics[width=0.45\textwidth]{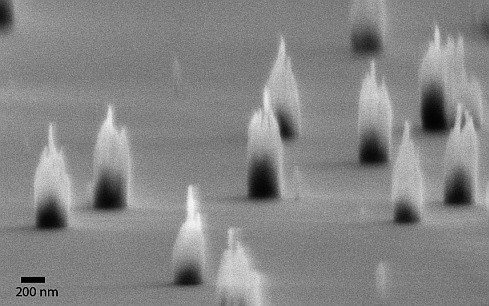} \\
 (e) & (f)\\
\includegraphics[width=0.45\textwidth]{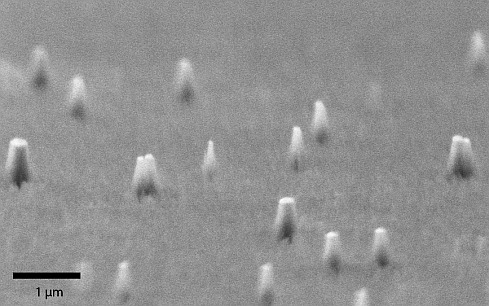} &
\hspace{0cm}\includegraphics[width=0.45\textwidth]{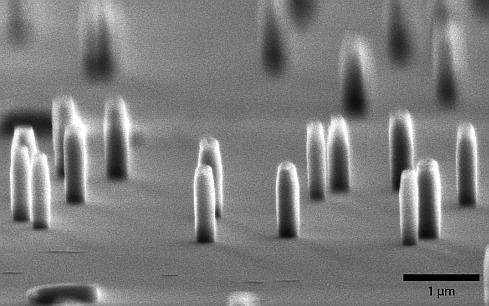}\\
 (g) & (h)\\
\end{array}$
\caption{Summary of etch recipe variations in DOI after a $\unit[5]{min}$ oxygen dry etch, using $\unit[250]{nm}$ $\mathrm{Au}$ particles as mask. Labels (a) through (h) correspond to etch recipes with the same name in Table \ref{tab:recvar}.}
\label{fig:recvar}
\end{figure*}

Based on our design presented earlier, an etch recipe and etch mask that yield wires with vertical profile and high selectivity are needed. Several dry etch parameters were varied in order to obtain an acceptable wire profile. The results are presented in Figure \ref{fig:recvar} and the etch parameter variations, heights and diameters are summarized in Table \ref{tab:recvar}. Waist diameter refers to the thickest part of the nanowire and the taper angle is calculated based on the waist and the bottom diameter.
The tabulated variations are for a $\unit[5]{min}$ etch using $\mathrm{Au}$ colloids with a diameter of $\unit[\approx250]{nm}$ as an etch mask on poly-D samples. A height of $\unit[1]{\mu m}$ was determined in most cases, except for (d) and (g) the height was $\unit[1.1]{\mu m}$ and $\unit[0.7]{\mu m}$ respectively.

\begin{table*}[bthp]
\caption{Summary of all etch variations. Straight wires were obtained with the optimized etch recipe (last row) for poly-D.}
\label{tab:recvar}
\begin{tabularx}{1\textwidth}{ccccccccccclX}
\toprule
Rec. & Temp. & Press. & Bias & ICP & $O_2$ & etch rate & Bottom $\diameter$ & Top $\diameter$ & Waist $\diameter$ & Taper angle & Result\\
 & $\mathrm{\,^{\circ} C}$ & $\mathrm{mTorr}$ & $\mathrm{W}$ & $\mathrm{W}$ & $\mathrm{sccm}$ & $\mathrm{nm/min}$ & $\mathrm{nm}$ & $\mathrm{nm}$ & $\mathrm{nm}$ & $\mathrm{\,^{\circ}}$ & \\
\cmidrule(r){1-11}\cmidrule(l){12-12}
(a) & RT & 20 & 100 & 700 & 30 & 200 & 330-420 & 270 & 330 & - & grass\\
(b) & 73 & 10 & 100 & 700 & 30 & 200 & 290 & 240 & 330 & 88 & undercut\\
(c) & 132 & 10 & 100 & 700 & 30 & - & - & 250 & 250 & - & grass\\
(d) & RT & 10 & 100 & 700 & 45 & 220 & 280 & 220 & 300 & 89.5 & straight\\
(e) & RT & 10 & 100 & 1000 & 30 & 200 & 220 & 220 & 310 & 87.4 & undercut\\
(f) & RT & 3 & 200 & 700 & 30 & - & 280 & - & 280 & - & erosion\\
(g) & RT & 20 & 200 & 700 & 30 & 140 & - & 200 & - & - & over cut\\
(h) & RT & 10 & 100 & 700 & 30 & 200 & 260 & 260 & 280 & 89.5 & $\approx$ straight\\
\bottomrule
\end{tabularx}
\end{table*}

Recipe (a) resulted in unwanted grass. Raising the temperature and decreasing the pressure resulted in smooth bottom surfaces (no grass) but introduced a noticeable mask erosion at the top (b). Further increase in temperature resulted in re-appearance of the grass (c) and a reduced etch depth. Slightly undercut wires are obtained with a higher oxygen flow rate of $\unit[45]{sccm}$ (recipe d) and a reduced temperature compared to recipe (b) as shown in Figure \ref{fig:recvar}d.  The increased flow rate and reduced temperature also increased the mask erosion at the top of the nanowires. To consider the effect of an increased ion density the ICP power was increased by 30\% (recipe (e)) while reducing the oxygen flow rate compared to recipe (d) by 33\%. This yielded in a 3.5 times larger undercut compared to recipe (d).
In recipe (f) (Figure \ref{fig:recvar}f) the chamber pressure was decreased to $\unit[3]{mTorr}$, the bias power was increased to $\unit[200]{W}$ and the oxygen flow rate was reduced to $\unit[30]{sccm}$ compared to recipe (d) which lead to mask erosion and excessive etching.
In Figure \ref{fig:recvar}g the etch recipe with $\unit[20]{mTorr}$ and $\unit[200]{W}$ was tested to minimize mask erosion by increasing the pressure compared to recipe (f). This resulted in relatively little mask erosion, an over cut wire profile and a reduced etch depth. To minimize the mask erosion observed in recipe (b) and (d) (Figures \ref{fig:recvar}b, \ref{fig:recvar}d), the oxygen flow rate was reduced to $\unit[30]{sccm}$ compared to recipe (d) as tabulated in recipe (h). This gave a near vertical wire profile with minimal mask erosion as shown in Figure \ref{fig:recvar}h and was used as the basic recipe for etching all the diamond samples presented in this paper.

\subsection{Comparison of nanoparticle etch mask}
\begin{figure}[bthp]
\begin{centering}
\subfloat[]{\begin{centering}
\includegraphics[width=0.29\columnwidth]{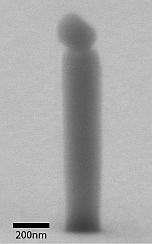}
\label{fig:al2o3}
\par\end{centering}
}\subfloat[]{\begin{centering}
\includegraphics[width=0.29\columnwidth]{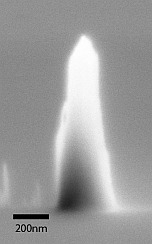}
\label{fig:si02}
\par\end{centering}
}\subfloat[]{\begin{centering}
\includegraphics[width=0.29\columnwidth]{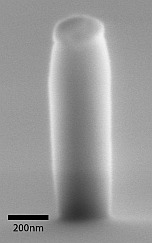}
\label{fig:au250_1}
\par\end{centering}
}
\par\end{centering}

\caption{Poly-D nanowires obtained with (a) $\mathrm{Al_2O_3}$ powder ($\unit[\approx200]{nm}$ in diameter), (b) $\mathrm{SiO_{2}}$ particles ($\unit[210]{nm}$ in diameter) and (c) $\unit[250]{nm}$ gold particles as mask, etched with the oxygen recipe. (a) The height is $\unit[\approx1]{\mu m}$ and the diameter is $\unit[200]{nm}$ both on the top and the bottom. The thickest diameter is $\unit[230]{nm}$. The particle height is $\unit[\approx210]{nm}$ and its thickest diameter is $\unit[230]{nm}$.
(b) The height of the wires is $\unit[1]{\mu m}$ and the diameter varies from $\unit[\approx342]{nm}$ on the bottom to $\unit[\approx20]{nm}$ at the top. (c) The nanowire is $\unit[\approx900]{nm}$ high. The diameter varied from $\unit[275]{nm}$ on the bottom, to a maximum of $\unit[310]{nm}$ and $\unit[250]{nm}$ at the top. The particle measures $\unit[245]{nm}$ in diameter and $\unit[130]{nm}$ in height. The etch time was $\unit[5]{min}$.}
\label{al2o3sio2}
\end{figure}

In order to improve the etch selectivity, different nanoparticle masks were also evaluated. A close up view of poly-D wires obtained using the optimized etch recipe with different nanoparticle masks are shown in Figure \ref{al2o3sio2}.

The $\mathrm{Al_2O_3}$ mask was tried first. A similar mask has been used successfully to build micropost-cavities with embedded quantum dots in the past \cite{Vuckovic+Santori}. $\mathrm{Al_2O_3}$ nanoparticles were deposited from both suspension and powder form. In the latter case, the samples were dipped top-down into the powder. $\mathrm{Al_2O_3}$ was found to be an ideal etch mask with no appreciable erosion.
A $\unit[4]{min}$ oxygen etch gave a height of the etched diamond structure of $\unit[\approx1]{\mu m}$ and a diameter of $\unit[200]{nm}$ both at the top and the bottom, as shown in Figure \ref{al2o3sio2}a. The thickest diameter was $\unit[230]{nm}$. The particle height was $\unit[\approx210]{nm}$ and its thickest diameter was $\unit[230]{nm}$. However, $\mathrm{Al_2O_3}$ nanoparticles were found to be clumping together, resulting in large etched features with random cross-sectional profiles (Figure \ref{fig:al2o3}), which was not desirable. Better uniformity and isolation of individual $\mathrm{Al_2O_3}$ nanoparticles can be achieved by modifying the $\mathrm{Al_2O_3}$ nanoparticles deposition process.

Due to better size and shape uniformity $\mathrm{SiO_{2}}$ nanoparticles were also evaluated. Figure \ref{fig:si02} shows the etch results obtained using individual $\mathrm{SiO_{2}}$ nanoparticles. Poor selectivity and a high mask erosion were observed, since $\mathrm{SiO_{2}}$ particles were etched at a rate of $\unit[\geq40]{nm/min}$. The mask erosion also resulted in a tapered nanowire profile. The height of the wires after a $\unit[5]{min}$ oxygen etch was $\unit[1]{\mu m}$, and the diameter varies from $\unit[\approx342]{nm}$ on the bottom to $\unit[\approx20]{nm}$ at the top. The $\mathrm{SiO_{2}}$ particles did not survive the etch.

Figure \ref{fig:au250_1} shows a $\unit[\approx900]{nm}$ high nanowire obtained with $\unit[250]{nm}$ gold particles as the mask. The diameter varied from $\unit[275]{nm}$ on the bottom to a maximum of $\unit[310]{nm}$ and $\unit[250]{nm}$ at the top. The particle shown in this image measures $\unit[245]{nm}$ in diameter and $\unit[130]{nm}$ in height. The etch time was $\unit[5]{min}$.

\begin{figure}[bthp]
\begin{centering}
\subfloat[]{\begin{centering}
\includegraphics[width=1\columnwidth]{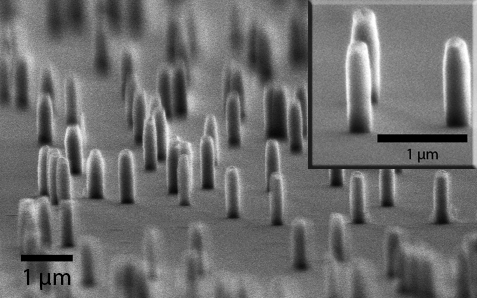}
\par\end{centering}
\label{fig:au250}}

\subfloat[]{\begin{centering}
\includegraphics[width=1\columnwidth]{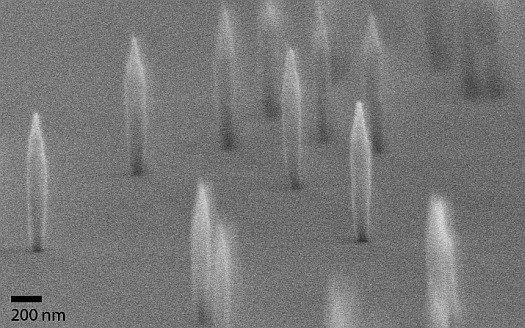}
\par\end{centering}
\label{fig:au100a}}
\end {centering}

\caption{Poly-D nanowires after a $\unit[5]{min}$ oxygen etch. (a) $\unit[250]{nm}$-Au colloids served as etch mask. The top-bottom height of the wires is $\unit[\approx1]{\mu m}$ and the thickest diameter is about $\unit[280]{nm}$. The diameter at the bottom of the wires is $\unit[\approx260]{nm}$. Inset: Close-up view of some wires. Figure (b) shows poly-D nanowires obtained with a 1:200 dilution of $\unit[100]{nm}$-$\mathrm{Au}$ colloids as a mask. The top-bottom height of the wires is $\unit[\approx0.9]{\mu m}$ and the thickest diameter is about $\unit[120]{nm}$.}
\label{fig:au250100}
\end{figure}

The Au nanoparticles were the easiest to disperse, resulting in single nanoparticles requiring no further processing stjpg. Using $\unit[250]{nm}$ nanoparticles, a $\unit[5]{min}$ etch resulted in $\unit[1]{\mu m}$ tall nanowires, as shown in Figure \ref{fig:au250}. The diameter varied from $\unit[260]{nm}$ at the bottom and top to $\unit[280]{nm}$ as thickest diameter. The etch rates of the mask and the DOI film were determined to be $\unit[25]{nm/min}$ and $\unit[200]{nm/min}$, respectively. This results in an etch selectivity of 8:1. The wires created in poly-D have a near-vertical profile. Figure \ref{fig:au250} and Figure \ref{fig:au100a} show a comparison of poly-D wires etched using $\unit[250]{nm}$ and $\unit[100]{nm}$ nanoparticles. The use of $\unit[100]{nm}$ Au particles (1:200 dilution) resulted in similar etch rates but with a tapered nanowire profile, which can be attributed to significant mask erosion in the case of the small gold nanoparticles (Figure \ref{fig:au100a}). However, such tapered profile may be beneficial for achieving high extraction efficiency for photons emitted from NV centers embedded within nanowires \cite{Lalanne}, \cite{Mork}. The height was $\unit[\approx0.9]{\mu m}$ and the thickest diameter is about $\unit[120]{nm}$.

\subsection{Single crystal diamond nanowires}\label{sec:sc-d}

\begin{figure}[bthp]
\subfloat[]{\begin{centering}
\includegraphics[width=1\columnwidth]{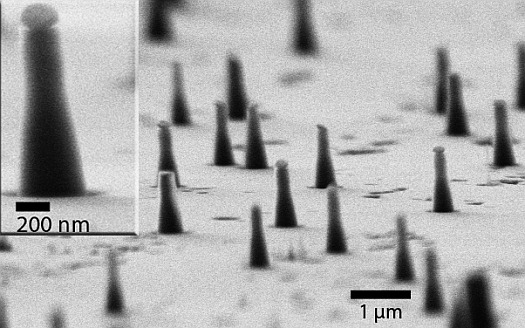}
\par\end{centering}

\label{fig:scda}}

\subfloat[]{\begin{centering}
\includegraphics[width=1\columnwidth]{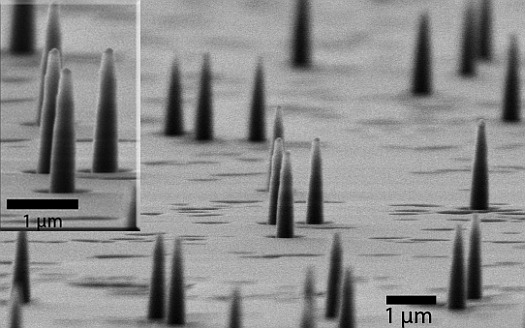}
\par\end{centering}
\label{fig:scdb}}

\caption{Sc-D nanowires obtained with $\unit[250]{nm}$-$\mathrm{Au}$ colloids as a mask for oxygen etching. (a) E6 IIa sc-D nanowires with a top-bottom wire height of $\unit[\approx1.1]{\mu m}$. The diameters vary from $\unit[\approx350]{nm}$ at the bottom of the wires to $\unit[\approx230]{nm}$ at the top. The etching time was $\unit[5]{min}$. Inset: Close-up view of one wire. (b) E6 Ib nanowires with a top-bottom height of $\unit[\approx1.9]{\mu m}$.
The diameters vary from $\unit[\approx470]{nm}$ at the bottom of the wires to $\unit[\approx190]{nm}$ at the top beneath the remaining particles. The etching time was $\unit[10]{min}$. Inset: Magnified view of three wires.}
\label{fig:scd}
\end{figure}

Due to the ease in dispersing and the reasonable etch selectivity of 8:1, $\mathrm{Au}$ nanoparticles were drop-casted on E6 CVD IIa and E6 HPHT Ib samples. The samples were etched for $\unit[5]{min}$ and $\unit[10]{min}$, respectively, the results of which are shown in Figure \ref{fig:scd}. After a $\unit[5]{min}$ etch, the nanowires are broadened in the bottom half and are almost straight in the upper half, as indicated in Figure \ref{fig:scda}. The top-bottom wire height is $\unit[\approx1.1]{\mu m}$. The diameters vary from $\unit[\approx350]{nm}$ at the bottom of the wires to $\unit[\approx230]{nm}$ at the top just beneath the remaining particles. The etching time was $\unit[5]{min}$. An etch rate of $\unit[220]{nm/min}$ was measured. Ib sc-D nanowires after a $\unit[10]{min}$ etch are shown in Figure \ref{fig:scdb}. The gold particle mask was almost completely eroded after this time and thus the upper part of the nanowire comes out tapered. The etch rate is $\unit[190]{nm/min}$. The top-bottom height is $\unit[\approx1.9]{\mu m}$ and the diameters vary from $\unit[\approx470]{nm}$ at the bottom of the wires to $\unit[\approx190]{nm}$ at the top beneath the remaining particles. The mask etch rate was in both cases $\unit[20-25]{nm/min}$. Thus the etch selectivity was found to be $\approx 8:1$ and $\approx 9:1$ for Ib and IIa sc-Ds respectively.

While the nanoparticle drop-casting approach provides a simple method to realize a large number of randomly positioned diamond nanowires, ordered arrays of nanowires are desirable for the characterization of functional devices. Another factor to consider is the effect of the near-spherical shape of the nanoparticle etch mask on the etch profile of nanowires. Due to these reasons, EBL-defined masks have also been considered in this work.

Ordered arrays of varying diameters between $\unit[100]{nm}$ and $\unit[250]{nm}$, which covers the range of the optimum diameters as presented in Section 2, were patterned on sc-D samples. Wire-to-wire distances (periodicity) of $\unit[2]{\mu m}$ and $\unit[3]{\mu m}$ were used. The distance was chosen to allow ease in characterization of individual nanowires using a custom built confocal micro- photoluminescence setup without introducing any stray signal from neighboring nanowires. Details of the experimental setup and results will be presented in a future publication.

\begin{figure}[bthp]
\begin{centering}
\includegraphics[width=1\columnwidth]{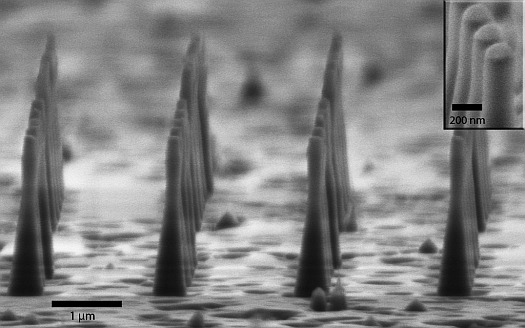}
\par\end{centering}

\caption{Sc-D nanowires obtained in a IIa CVD diamond. After a $\unit[10]{min}$ oxygen etch using an EBL defined evaporated Cr/Au (10/200)nm mask.
The height is $\unit[2.2]{\mu m}$ from top to bottom, the remaining mask on this set of wires is $\unit[\approx60]{nm}$ high. The diameter just beneath the mask is $\unit[\approx160]{nm}$ and at the bottom $\unit[\approx490]{nm}$.
Inset: The etch time was $\unit[5]{min}$, the remaining mask on this set of wires is $\unit[\approx120]{nm}$ high. The diameter just beneath the mask is $\unit[\approx200]{nm}$.
}

\label{fig:scdpmma}
\end{figure}

EBL using PMMA resists followed by a lift-off process was used to realize an evaporated Cr/Au metal mask, as described in Section 3 (Figure \ref{fig:cartoon}). The results of etching E6 IIa CVD diamond with such a mask are shown in Figure \ref{fig:scdpmma}. A $\unit[10]{min}$ oxygen etch produces $\unit[2.2]{\mu m}$ high nanowires. The diameter at the top of the wires of this array is $\unit[\approx160]{nm}$ and at the bottom $\unit[\approx490]{nm}$. The diamond etch rate is $\unit[220]{nm/min}$, whereas the etch rate of the evaporated gold is $\unit[26]{nm/min}$. In the inset a wire is shown with the remaining mask after a $\unit[5]{min}$ etch. The mask is $\unit[\approx120]{nm}$ high and the diameter of the wire is $\unit[\approx200]{nm}$ at the top. Again, an over cut etch profile is observed in the case of sc-D nanowires. As mentioned above, the same recipe resulted in a vertical nanowire profile in the case of poly-D nanowires. We expect that an increase in the isotropic etch component of our recipe would result in straighter nanowires.

\begin{figure}[bthp]
\begin{centering}

\subfloat[]{\begin{centering}
\includegraphics[width=1\columnwidth]{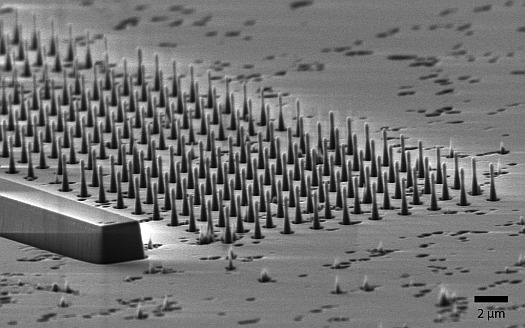}
\par\end{centering}
\label{fig:scdFOxa}}

\subfloat[]{\begin{centering}
\includegraphics[width=0.48\columnwidth]{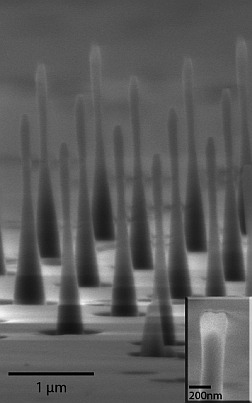}
\par\end{centering}
\label{fig:scdFOxb}}
\subfloat[]{\begin{centering}
\includegraphics[width=0.48\columnwidth]{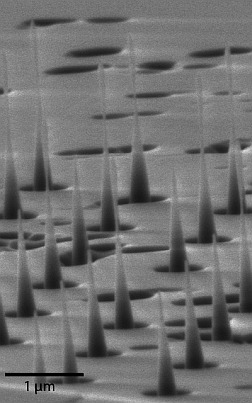}
\par\end{centering}
\label{fig:scdFOxc}}

\par\end{centering}

\caption{Sc-D wires formed in a Ib HPHT diamond. FOx was used to produce an array of pillar-shaped mask with EBL. The mask was then transfered to the substrate during a $\unit[10]{min}$ oxygen etch and subsequently removed.
(a) One pattern of wires with the same diameters. (b) The height of the wires is $\unit[2.3]{\mu m}$ from top to bottom. The diameters at the broadest part of the top and the bottom are $\unit[\approx110]{nm}$ and $\unit[\approx310]{nm}$ respectively. At the thinnest part of the wires the diameter is $\unit[\approx70]{nm}$. Inset: One wire with the remaining FOx mask after etching (different diameter). (c) One array of $\unit[2.2]{\mu m}$ high nanowires with a bottom diameter of $\unit[226]{nm}$.}

\label{fig:scdFOx}
\end{figure}

$\mathrm{Au}$ nanoparticles and evaporated $\mathrm{Au}$ gave a similar etch rate of $\unit[25]{nm/min}$. In order to further simplify the fabrication procedure, we used FOx (which is a negative e-beam resist) as an etch mask, as mentioned in Section 3 (Figure \ref{fig:cartoon}). This required no metal evaporation and lift-off. The profile of the wires in E6 Ib sc-D can be seen in Figure \ref{fig:scdFOx}. Figure \ref{fig:scdFOxa} shows $\unit[2.3]{\mu m}$ high wires. The EBL-written diameter of this array was $\unit[150]{nm}$. The measured diameters at the broadest part of the top and the bottom are $\unit[\approx110]{nm}$ and $\unit[\approx310]{nm}$, respectively. At the thinnest part of the wires the diameter is $\unit[\approx70]{nm}$. In Figure \ref{fig:scdFOxc}, an array of $\unit[2.2]{\mu m}$ high nanowires with a bottom diameter of $\unit[226]{nm}$ are shown (written diameter: $\unit[100]{nm}$). We found the etch rate of FOx to be smaller than $\unit[10]{nm/min}$, which is almost three times better than the metal mask.

\begin{figure}[bthp]

\begin{centering}
\subfloat[]{\begin{centering}
\includegraphics[width=0.8\columnwidth]{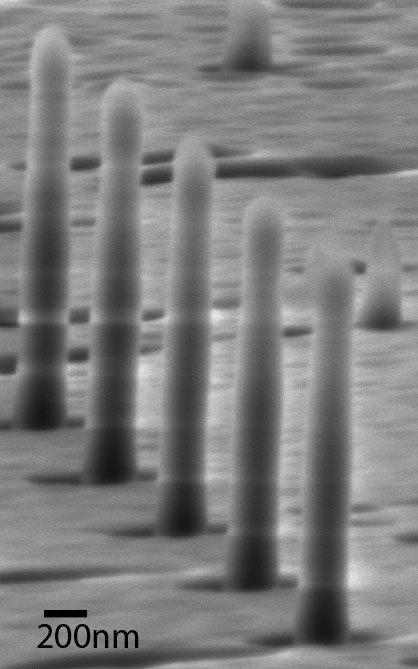}
\par\end{centering}
\label{fig:scdFOx2}}

\subfloat[]{\begin{centering}
\includegraphics[width=0.48\columnwidth]{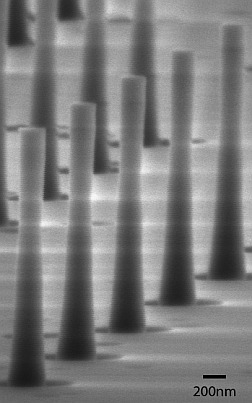}
\par\end{centering}
\label{fig:apollob}}
\subfloat[]{\begin{centering}
\includegraphics[width=0.48\columnwidth]{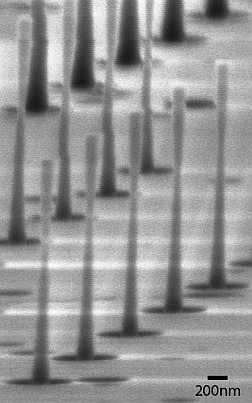}
\par\end{centering}
\label{fig:apolloc}}

\par\end{centering}

\caption{(a) Sc-D (Ib HPHT) wires after a $\unit[10]{min}$ oxygen etch. The FOx mask was removed.
The height of the wires is $\unit[1.9]{\mu m}$ from top to bottom and the diameter at the bottom is  $\unit[\approx260]{nm}$. (b) and (c) Sc-D nanowires formed in a diamond from Apollo Diamond after removing the FOx mask. Their height is $\unit[2.4]{\mu m}$ and the distance between each nanowire is $\unit[3]{\mu m}$. (b) Nanowires with a bottom diameter of $\unit[340]{nm}$. The thinnest and top diameters are $\unit[190]{nm}$ and $\unit[210]{nm}$ respectively. (c) The diameter varies from $\unit[170]{nm}$ at the bottom, to $\unit[60]{nm}$ at the nanowire waist and $\unit[120]{nm}$ at the top.}

\end{figure}

In order to improve the etch profile of sc-D nanowires and obtain nanowires with vertical sidewalls, we changed our etch recipe slightly by varying the ICP power. Based on the profile we obtained with our base etch recipe, we decided to decrease the chemical etch rate after a few minutes and then increase it to prevent the broadening of the wires (Figures \ref{fig:scdpmma} and \ref{fig:scdFOx}). We used our usual recipe for $\unit[2]{min}$, decreased the ICP power to $\unit[600]{W}$ for $\unit[3]{min}$ and then ramped up to $\unit[1000]{W}$ for $\unit[5]{min}$. The results can be seen in Figure \ref{fig:scdFOx2}. The periodicity and diameter of the wires used in EBL are $\unit[3]{\mu m}$ and $\unit[200]{nm}$ respectively. We were able to realize $\unit[1.9]{\mu m}$ tall nanowires with near vertical profile (bottom diameter of $\unit[\approx260]{nm}$) and a tapered top due to the high ICP power causing mask erosion. This can be minimized by using a thicker FOx mask film. As mentioned earlier, however, a tapered profile may be preferable \cite{Lalanne} for increasing the extraction collection efficiency from emitters embedded inside nanowires.

Based on the above recipe optimization results, an ultra pure type IIa diamond from Apollo Diamond, Inc. was also etched with slightly different etch step lengths: $\unit[1.5]{min}$, $\unit[3]{min}$ and $\unit[5.5]{min}$. The profile of the $\unit[2.4]{\mu m}$ tall wires with a periodicity of $\unit[3]{\mu m}$ can be seen in Figures \ref{fig:apollob} and \ref{fig:apolloc}. Nanowires with a bottom diameter of $\unit[340]{nm}$ are shown in Figure \ref{fig:apollob} (EBL written diameter: $\unit[220]{nm}$). The thinnest and top diameters are $\unit[190]{nm}$ and $\unit[210]{\mu m}$ respectively. Another array with diameters ranging from $\unit[170]{nm}$ at the bottom to $\unit[60]{nm}$ at the nanowire waist and $\unit[120]{nm}$ at the top can be seen in Figure \ref{fig:apolloc}. The diameter used in EBL was $\unit[140]{nm}$. The wire etch rate was measured as $\unit[240]{nm/min}$.

\begin{table}[bthp]
\caption{Summary of the etch rates in the different diamond types. For the Apollo IIa a different etch recipe has been used (see section \ref{sec:sc-d}), for the other diamond types we used the etch recipe introduced in section \ref{matandmet}.}
\label{tab:rates}
\begin{tabularx}{1\columnwidth}{lXcc}
\toprule
Type & Etch rate & etch time \\
 & $\mathrm{nm/min}$ & $\mathrm{min}$ \\
\cmidrule(r){1-3}
poly-D & 200 & 5 \\
Ib sc-D & 190 & 10 \\
IIa sc-D (E6) & 220 & 5 \\
IIa sc-D (Apollo Diamond) & 240 & 10 \\
\bottomrule
\end{tabularx}
\end{table}

\section{Conclusion}
In this work, we have presented the design and fabrication of diamond nanowires. Our modeling shows an optimal range of diameters for coupling (>80\%) the Nitrogen Vacancy (NV) center emission to the propagating waveguide mode inside the diamond nanowire, and collecting that emission with a high extraction efficiency using a 0.95 NA objective.
 We have also investigated diamond etching of poly-D and sc-D nanowires. 
We have summarized the etch rates of all diamond types in table \ref{tab:rates}.
 Nanowires with near vertical profiles were fabricated using an oxygen ICP RIE recipe. $\mathrm{Al_2O_3}$, $\mathrm{Au}$ and $\mathrm{SiO_2}$ particles, evaporated $\mathrm{Au}$ and FOx e-beam resist were evaluated as an etch mask. It was found that $\mathrm{Al_2O_3}$ nanoparticles were most resistant to etching. $\mathrm{Au}$ nanoparticles and evaporated $\mathrm{Au}$ gave a similar etch rate, while FOx e-beam resist gave an etch rate of $\unit[10]{nm/min}$ and was found to be the most suitable mask based on our fabrication process. The crystalline nature of the samples affected the etch profile and required different etch parameters to obtain nearly vertical etch profiles of the nanowires. Polycrystalline and type Ib single crystal diamond were found to etch at $\unit[(190-200)]{nm/min}$, whereas type IIa single crystal diamonds were found to etch at about $\unit[(220-240)]{nm/min}$. We believe that our nanowire work and the diamond nanophotonic technology that we are developing will play an important role as an enabling platform for fundamental quantum information processing applications, as well as bio-sensing applications based on magnetometry \cite{Maze}.

\section{Acknowledgments}
This work is supported in part by the DARPA Quest program (PM Jagdeep Shah), NSF NIRT grant, and Harvard NSEC. The authors wish to acknowledge the help during the search for a suitable nanowire fabrication process by Chih-Hsun Hsu and Jeffrey Shainline (Jimmy Xu's lab at Brown University). The authors thank Advanced Diamond Technologies and Apollo Diamond for providing diamond test samples. Fruitful discussions with Philippe Lalanne and help from Maja Cassidy and Ben Hatton are acknowledged.
One of the authors (B. H.) wishes to cordially thank Anna Fontcuberta i Morral and Jonathan Finley for their support.  Most of the fabrication work was performed at the Center for Nanoscale Sciences (CNS) Nanofabrication facility at Harvard University, , a member of the National Nanotechnology Infrastructure Network (NNIN), which is supported by the National Science Foundation under NSF award no. ECS-0335765. Disclaimer: The views, opinions, and/or findings contained in this publication are those of the authors and should not be interpreted as representing the official views or policies, either expressed or implied, of the Defense Advanced Research Projects Agency or the Department of Defense..

\end{document}